\documentclass[letterpaper, 10pt, conference]{ieeeconf}

\IEEEoverridecommandlockouts

\usepackage[active]{srcltx}

\usepackage{verbatim}
\usepackage{epsfig}
\usepackage{amsmath}
\usepackage{amssymb}
\usepackage{psfrag}
\usepackage[T1]{fontenc}
\usepackage{graphicx}
\usepackage[usenames,dvipsnames]{pstricks}
\usepackage{pst-grad} 
\usepackage{pst-plot} 
\usepackage{algorithmic}

\graphicspath{{./figure/}}

\title{\LARGE \bf
Realizing Information Erasure in Finite Time}

\author{ James. Melbourne,  Saurav. Talukdar, Murti. V. Salapaka
\thanks{James Melbourne is with Electrical and Computer Engineering, University of Minnesota-Twin Cities, 
        {\tt\small melbo013@umn.edu}}%
\thanks{Murti V. Salapaka is with Electrical and Computer Engineering, University of Minnesota-Twin Cities, 
        {\tt\small @umn.edu}}%
\thanks{Saurav Talukdar is with Mechanical Engineering,University of Minnesota-Twin Cities, 
    {\tt \small @umn.edu}}
}

\begin{document}

\maketitle
\thispagestyle{empty}
\pagestyle{empty}

\begin{abstract}
In this article, we focus on erasure of a bit of information in finite time. Landauer's principle states that the average heat dissipation due to erasure of information is $k_BT\ln{2}$, which is achievable only in an asymptotic manner.  Recent theoretical developments in non-equilibrium thermodynamics and stochastic control, predict a more general bound for finite time erasure dependent on the Wasserstein distances between the initial and final configurations.  These predictions suggest improvements to experimental protocol with regards to minimizing average heat dissipation for bit erasure in finite time from a bistable well, under overdamped Langevin dynamics. We present a comparative study of a theoretically optimal protocol with an existing protocol, and highlight the closeness and deviation from optimality. 
\end{abstract}

\section{INTRODUCTION}
Improving computational efficiency with regard to energy consumption while reduction in physical size is a persistent theme of research for chip designers \cite{gammaitoni2015towards}. There is a strong push toward minimizing heat dissipation in phones and computers for their longevity and better user experience. An interesting question in this direction is, whether there is a fundamental limit to energy consumption in computations, which will eventually act as a bottleneck in this journey of improving computational efficiency. The knowledge of such fundamental limits will shed light on the scope of improvement in existing technologies as well as suggest possible mechanisms to operate at the fundamental energy limits.

To seek answers to the questions raised, we consider the basic computational unit of a single bit memory and computations associated with it. For example: the NOT operation on the information stored in a single bit memory, can be done without any associated heat dissipation \cite{feynman2000feynman}. Another important fundamental operation at the bit level, is the reset (also known as erasure) operation. Landauer's principle \cite{landauer1961irreversibility} gives a lower bound on the heat dissipation associated with the erasure of information stored in a single bit memory. It states that, erasure of information is necessarily dissipative and is accompanied by $k_BT\ln{2}$ ($\sim 10^{-21}J$) amount of average heat dissipation per erased bit. It is a fundamental result linking thermodynamics with information processing operations. The scale of the energy specified by the Landauer's bound necessitates the use of nano scale systems with pico-Newton force resolution. Landauer's principle was experimentally validated only recently (almost 50 years after being first proposed by Rolf Landauer) by several researchers \cite{berut2012experimental, jun2014high, talukdar2017memory, hong2016experimental} only after significant scientific advances in nanotechnology and nano scale probing. 

It is important to understand that information erasure with average dissipation equal to the Landauer's bound can only be achieved by a quasi static process.  
The above mentioned experimental studies focused solely on approaching the Landauer bound in an asymptotic manner using a quasi-static process. However, the erasure of a physical bit of information is always in a finite time $t$, and in practice small $t$ is generally desired.  In this article we focus on erasure mechanisms which are optimal with regards to the associated average heat dissipation for finite time processes. To determine the optimal erasure protocol(with respect to average heat dissipation) for finite time erasures, the authors rely on some of the most recent tools developed in steering of the distribution of a linear system from an initial to a final distribution \cite{chen2017optimal} and the connections of such tools with non equilibrium thermodynamics \cite{aurell2012refined,chen2018stochastic}.  

As we will see, these bounds (up to a constant $c$ dependent on the spacial, ``Wasserstein'' distance of the distributions associated to pre and post erasure) sharpen Landauer to $k_B T \ln{2} + c/t$ for an erasure over a time interval of length $t$.  One recovers Landauer in the asymptotic limit, while for time efficient erasures ($t \to 0$), the effects of the added term dominate the contribution to heat dissipation.  The study of achievable $c$ is outside the scope of this letter, we direct the reader to \cite{talukdar2018analyzing} (and further generalized in \cite{melbourne2018error, melbourne2018deficit}) where the authors have begun a study that addresses achievable scale for erasure.

In this article, we utilize an experimentally characterized Monte Carlo simulation framework developed by the authors \cite{bhaban2016noise, talukdar2017memory} to study optimal erasure protocols. This simulation framework and the erasure protocol in \cite{talukdar2017memory} (referred as the duty ratio protocol) is used to generate the initial and final probability distributions for a fixed time to accomplish erasure of information. Using, the obtained initial and final distributions, with the optimal mass transport tools from \cite{aurell2012refined, chen2018stochastic}, the authors compute the minimal cost of erasure in finite time. These lower bounds are compared with the cost of erasure in the author's prior work.  We highlight that for certain duty ratio protocol approaches the theorized minimal cost. Moreover, we hypothesize an improved protocol, whose intermediate steps of transporting the system from the initial to the final distribution traverse a Wasserstein-2 geodesic \cite{villani2008optimal}, which we describe explicitly. 

In Section II and III, we describe the physical system of interest, followed by a discussion on erasure and the duty ratio protocol in Section IV. In Section V, we review basic notions of thermodynamics followed by optimal erasures in Section VI. Optimal protocols are discussed in Section VII, followed by Results in Section VIII with Conclusions and Future Work in Section IX.

\section{A Brownian particle in a Laser Trap}

\begin{figure}[tb]
	\centering
	\includegraphics[scale = 0.35]{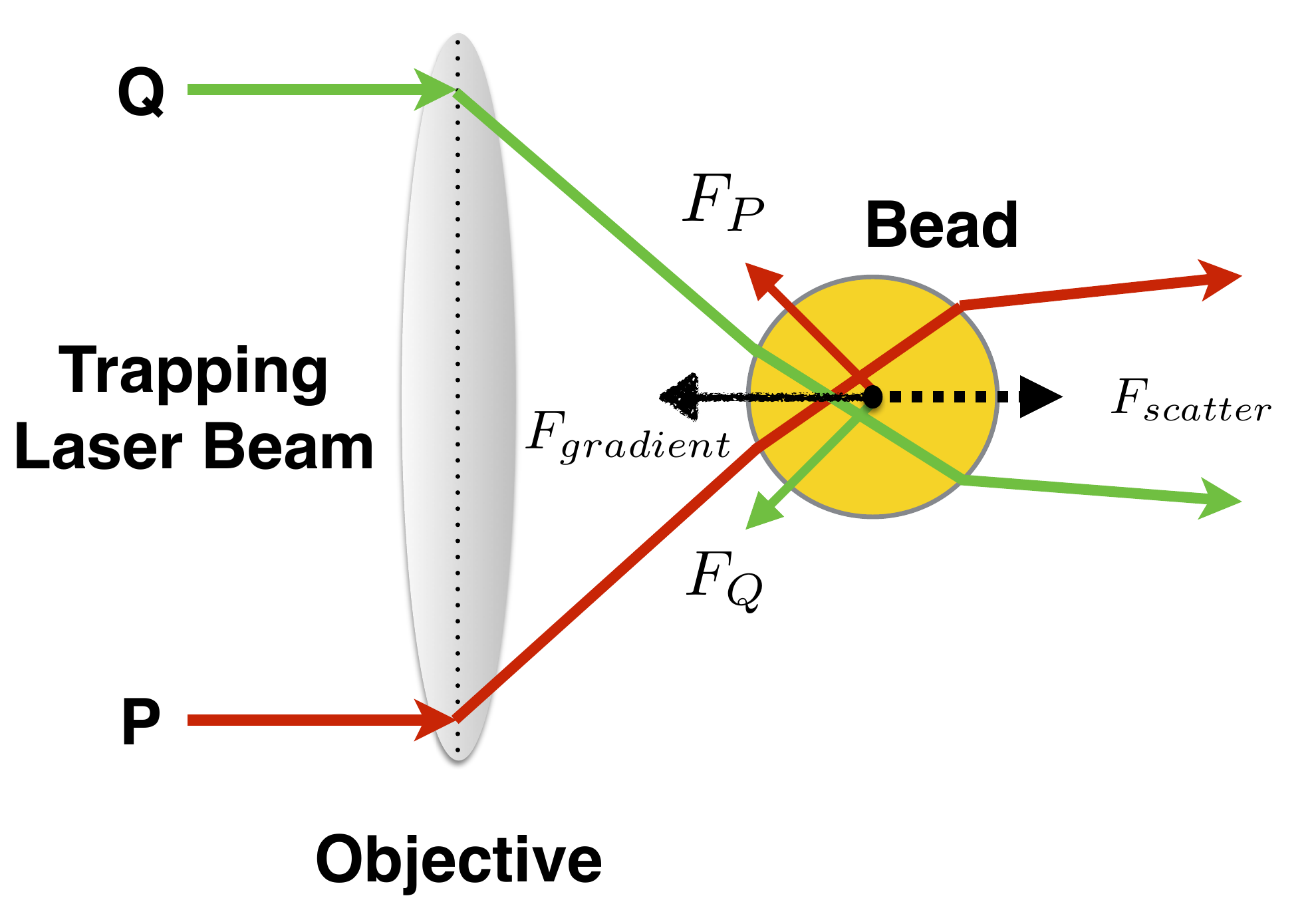} 
	\vspace{-0.1 cm}
	\caption{Bead in an optical trap}
	\label{fig:bead in trap}
\end{figure}

\begin{figure}[tb]
	\centering
	\includegraphics[scale = 0.27]{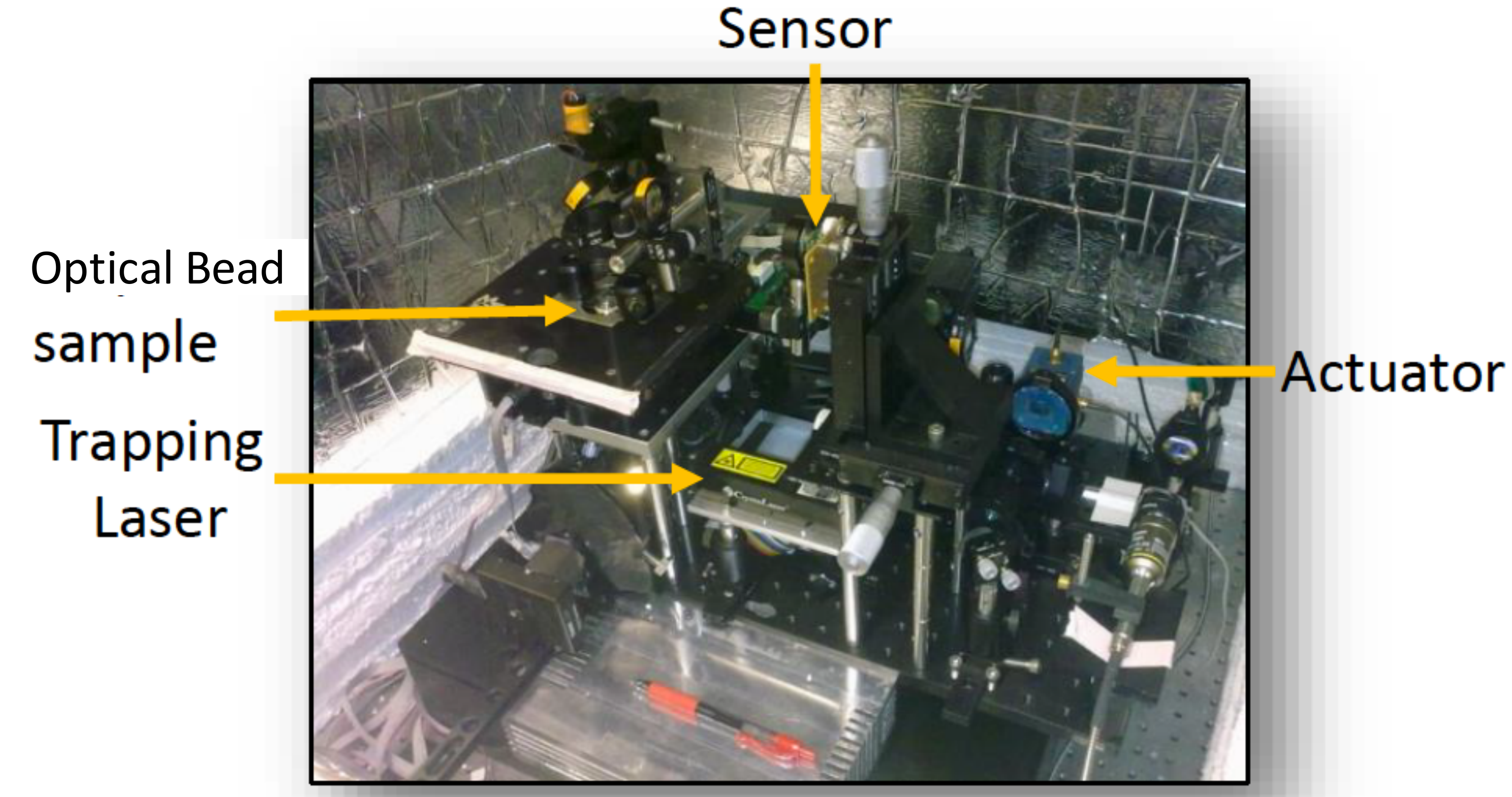} 
	\vspace{-0.1 cm}
	\caption{Optical tweezer setup.}
	\label{fig:tweezer}
\end{figure}

The physical system of interest comprises a Brownian particle in thermal equilibrium with a liquid medium at a constant temperature, $T$, under the influence of a potential, $U(x)$. 
Arthur Ashkin \cite{ashkin1986observation} demonstrated that when a laser beam is passed through an objective with high numerical aperture, and is incident on a spherical bead with appropriate refractive index, then, the momentum transfer from the reflected and refracted rays onto the bead results in a stable equilibrium point for the bead. Thus the opto-mechanical forces trap the bead. The restoring forces experienced by the optical bead in such a stable trap vary linearly for small displacements away from the equilibrium point, where the potential is of harmonic nature. An experimental realization of such a physical system is a spherical optical bead(Brownian particle) of diameter $\sim 1 \mu m$(micro meter) immersed in deionised water and under the influence of an optical trap(potential $U(x)$) created by a laser beam as shown in Figure \ref{fig:bead in trap}. The experimental realization of such a system is obtained in this study by a custom optical tweezer setup (see Figure~\ref{fig:tweezer}), where a bead with diameter 1~$\mu m$ is immersed in deionized water and is trapped optically. The trap  provides a potential $U(x)$ for the bead with location $x$ with respect to the trap and is excited by thermal noise. Thus the system of a Brownian particle in potential $U(x)$ is obtained.

\begin{figure}[tb]
	\centering
	\includegraphics[scale = 0.35]{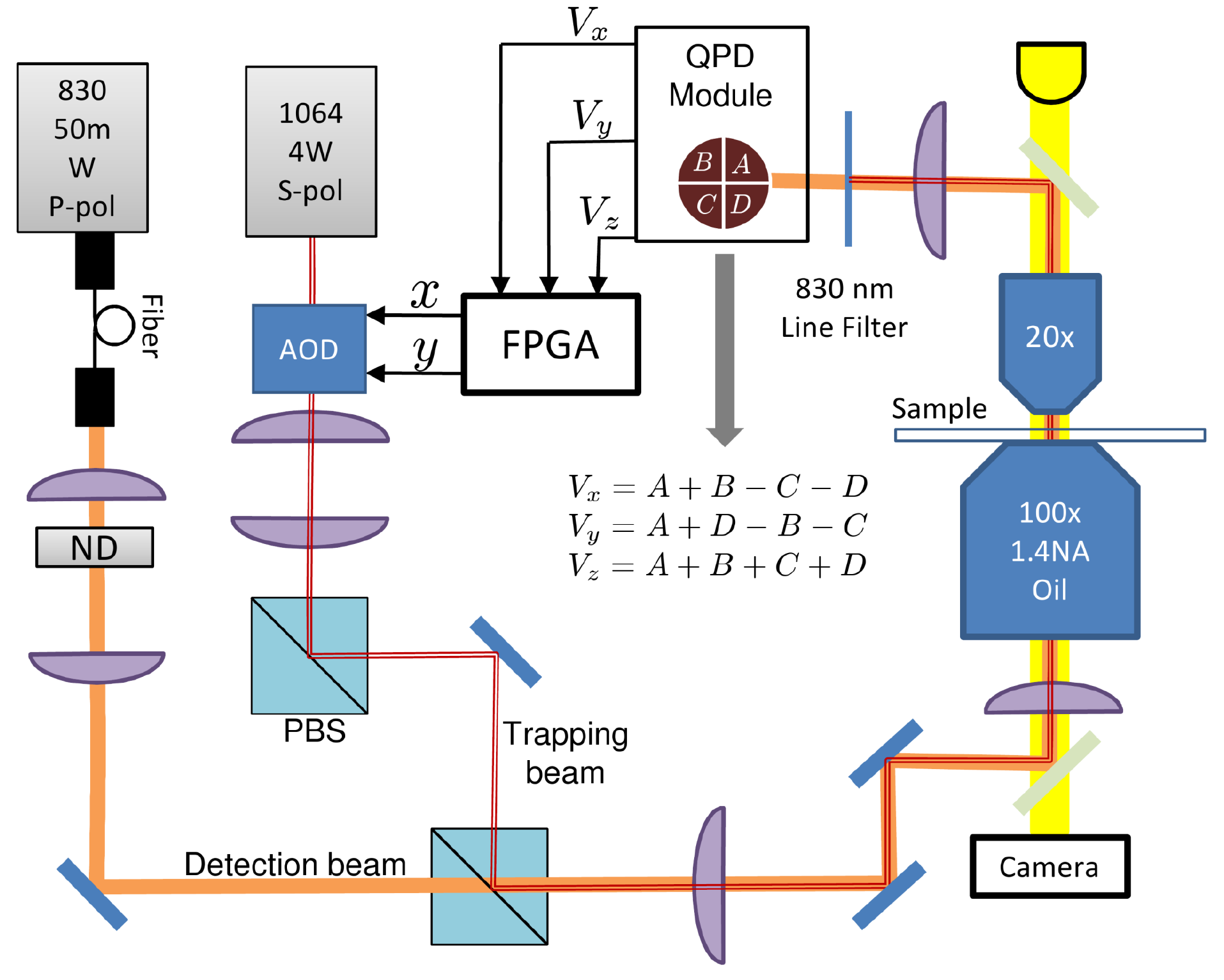} 
	\caption{Schematic for our Optical Tweezer setup.}
	\label{fig:schematic}
	\vspace{-0.7 cm}
\end{figure}

\begin{figure}
	\centering
	\includegraphics[scale = 0.41]{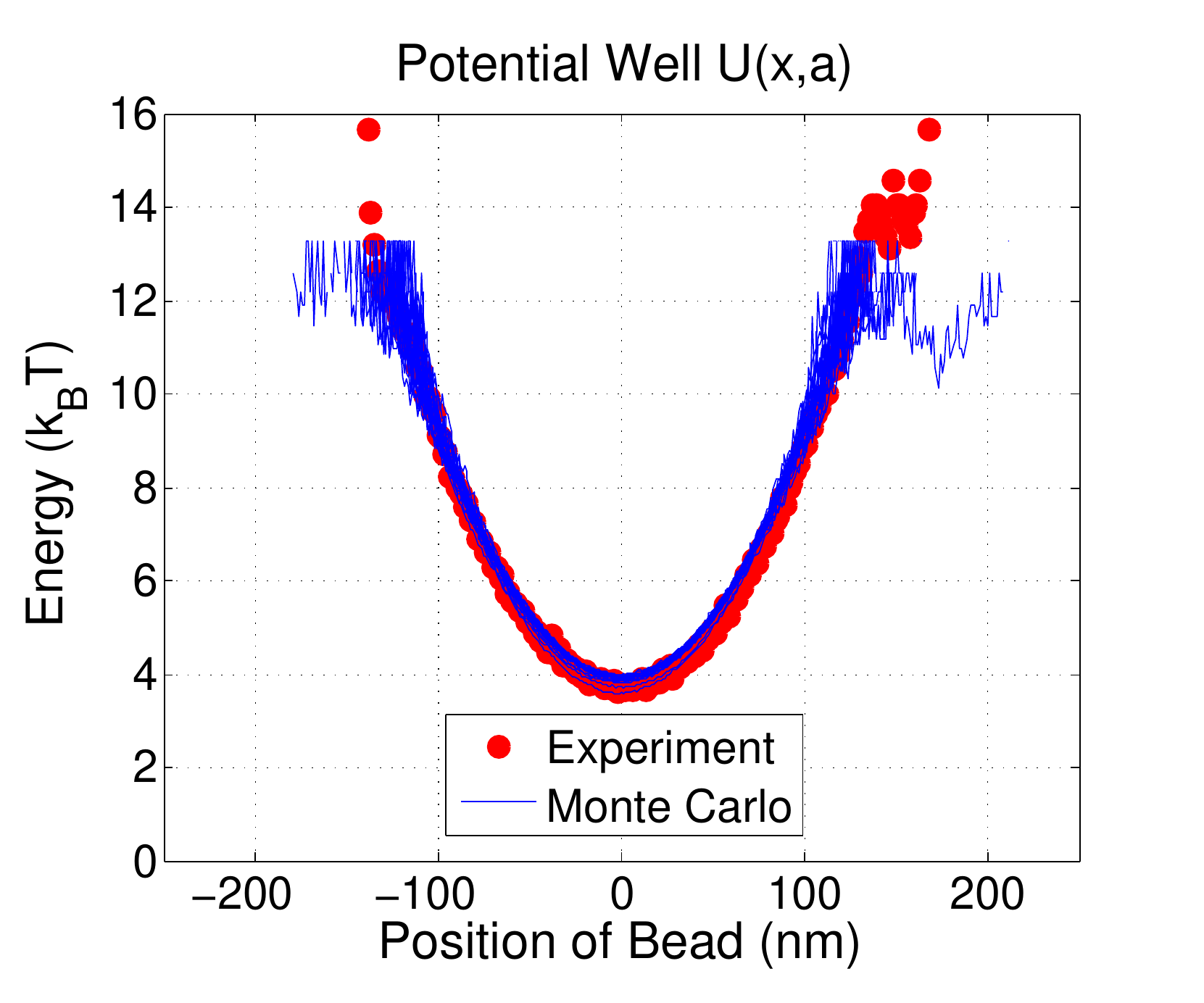}
	\vspace{-0.3 cm}
	\caption{Potential for single well obtained from Monte Carlo simulations and experiments. Experimental observations and Monte Carlo simulation predictions match well \cite{bhaban2016noise}.}
	\label{fig:single_well_pot}
\end{figure}

The experimental setup for creating an optical trap comprises of lasers and optics, actuators for manipulating trap position (using an Acusto Optic Deflector (AOD)), high resolution sensors (nano-meter level accuracy) for the measurement of position of the bead (using a quadrant photo diode(QPD)) and a Field Programmable Gate Array (FPGA) board for real time control.The schematic of the resulting optical tweezer setup is shown in Figure \ref{fig:schematic}. 
The position of the optical bead, $x(t)$, is measured for a sufficient duration ($100$ times the time constant of the bead dynamics) and binned to obtain the equilibrium probability distribution, $P(x)$, of the position of the bead. When the bead is under thermal equilibrium, the probability distribution,  $P(x)$, is given by the Boltzmann distribution,
\begin{align*}
P(x) = C ~\text{exp} ~(-\frac{U(x)}{k_BT}),
\end{align*}
where $C$ is a normalization constant. Thus the potential,
\begin{align}\label{eqn: potential}
U(x) = - k_B T ~\text{ln} ~(\frac{P(x)}{C}).
\end{align}
 $P(x)$ is obtained by binning experimentally measured bead position. Equation (\ref{eqn: potential}) is used to obtain $U(x)$ (see the red curve in Figure \ref{fig:single_well_pot}). 
It is important to note that the extent of experimentally obtained potential is finite, due to finite width of the laser beam. We now present a mathematical model of a particle in an optical trap, which can be used to perform Monte Carlo simulation studies. 

The dynamics of the bead in an optical trap is modeled by an overdamped Langevin equation \cite{gardiner1985handbook}. The discretized version of the dynamics is given by,
\begin{align}\label{eqn: discrete_langevin}
x(t+ dt) = x(t) -\frac{k}{\gamma}x(t)dt + \sqrt{\frac{2k_BT}{\gamma}dt}\ \nu(t), 
\end{align}
where, $k$ is the stiffness of the trap, $\gamma $ is the viscous friction coefficient, $k_B$ is the Boltzmann constant, $T$ is the temperature of the heat bath (liquid medium), $\nu(t) \in \mathcal{N}(0,1)$ with $\langle \nu(t)\nu(t')\rangle = \delta(t-t')$ and $dt$ is the time-step. Here, $\mathcal{N}(0,1)$ denotes the standard normal distribution, $\langle . \rangle$ denotes the expectation operator and $\delta(.)$ is the Dirac delta function. Outside the influence of the optical trap, the bead undergoes a random walk and does not experience any force from the potential $U(x)$; here the system dynamics is described by,
\begin{align}\label{eqn: discrete_brownian}
x(t+dt) = x(t) + \sqrt{\frac{2k_BT}{\gamma}dt}\ \nu(t).
\end{align}
Now we summarize the experimentally obtained values of the parameters of the model in (\ref{eqn: discrete_langevin}) and (\ref{eqn: discrete_brownian}). The trap stiffness $k = 0.0044~pN/nm$ is obtained from the bead position data using the Equipartition Theorem \cite{reif2009fundamentals},
\begin{align}\label{eqn: equipartition}
\frac{1}{2} k \langle x^2 \rangle = \frac{1}{2} k_B T,
\end{align}
where, the room temperature $T$ is $300 K$. Viscosity of deionised water, $\eta=8.9 \times 10^{-4} Pa.s$ \cite{huber2009new}. Thus, using Stoke's Law, $\gamma = 6\pi \eta r = 8.3 \times 10^{-9} Ns/m$, where, $r$, is the radius of the optical bead. The finite extent, $w$, of the potential $U(x)$, is determined to be $175$ nm. Thus, (\ref{eqn: discrete_langevin}) is employed when, when $|x(t)|< w$, and (\ref{eqn: discrete_brownian}) is employed when the bead position $x$ is in the regime $|x|>w$. The integration time step $dt$ is taken as $10^{-5} s$. The potential obtained by using (\ref{eqn: potential}) with equilibrium probability distributions from $100$ Monte Carlo simulations of 10 second duration (based on experimentally determined system parameters) using (\ref{eqn: discrete_langevin}), (\ref{eqn: discrete_brownian}) is shown in blue in Figure \ref{fig:single_well_pot}. A close match between experimental observation and the Monte Carlo simulations is observed, justifying the validity of the proposed first order model. In the next section,  we present a way to realize a Brownian particle in a bi-stable potential to model a single bit memory.

\section{A Brownian particle in a bi-stable well}
\begin{figure}
	\centering
	\includegraphics[scale = 0.35]{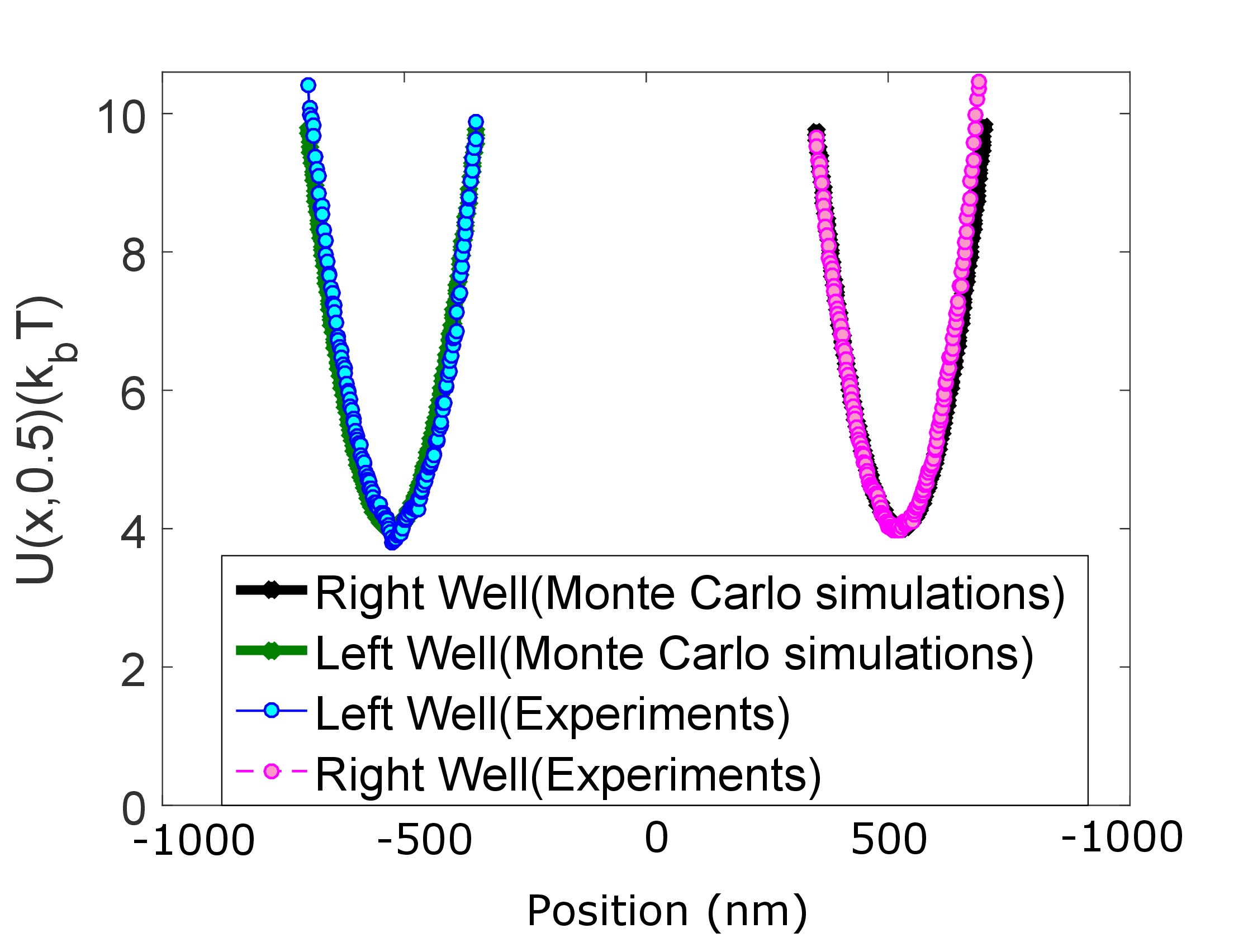} \vspace{-0.3 cm}
	\caption{Bi-stable potential for $L=550~nm$ obtained using Monte-Carlo simulations and experiments \cite{talukdar2017memory}. 
	}
	\label{fig:550nm_multiplex}
	\vspace{-0.4 cm}
\end{figure}

A bi-stable potential comprises two stable wells separated by a barrier.
It can be realized by multiplexing the trapping laser between two locations, $L$ and $-L$, alternately.  Here, the laser needs to be multiplexed an order faster than the time-scale of the bead dynamics. The bead effectively experiences a stable trap in either location. The dynamics of the bead is governed by the overdamped Langevin equation if it is under the influence of the laser trap at $L$ (see (\ref{eqn: discrete_double_well_langevin_d}) below) or $-L$ (see (\ref{eqn: discrete_double_well_langevin_-d}) below), otherwise it undergoes a random walk, where the particle dynamics is described by (\ref{eqn: discrete_double_well_brownian}). Thus the dynamics is given by, 
\begin{align}\label{eqn: discrete_double_well_langevin_d}
x(t+dt) &= x(t) -\frac{k}{\gamma}(x(t) - L)dt + \sqrt{\frac{2k_BT}{\gamma}dt}\ \nu(t),\\
\nonumber \text{if} \ s(t) &= 0 \  \text{and} \ |x - L| \leq w 
\end{align}
\begin{align}\label{eqn: discrete_double_well_langevin_-d}
x(t+dt) &= x(t) -\frac{k}{\gamma}(x(t) + L)dt + \sqrt{\frac{2k_BT}{\gamma}dt}\ \nu(t),\\ 
\nonumber \text{if} \ s(t) &= 1 \ \text{and} \ |x + L| \leq w 
\end{align}
\begin{align}\label{eqn: discrete_double_well_brownian}
x(t+dt) = x(t) + \sqrt{\frac{2k_BT}{\gamma}dt}\ \nu(t), \ \text{otherwise}.
\end{align}
Here, if laser is focused at $-L$, $s(t) = 1$, and if it is focused at $L$, then $s(t) = 0$. The laser moves periodically between $L$ and $-L$ with the laser residing for a duration at each location.  We define duty ratio, $d$ as the ratio of duration the laser spends at $-L$ over the total cycle time. The potential energy landscape obtained from 100 Monte Carlo simulations with $L= 550 nm$ for $30$ seconds each with a duty ratio of $0.5$ (equal time at both locations) using (\ref{eqn: potential}) is shown in Figure \ref{fig:550nm_multiplex}. We also implemented the same protocol in experiments and obtain the potential energy landscape using (\ref{eqn: potential}). Note that, the physical parameters derived for a single well in the previous section are used for the simulations for the bi-stable potential and no extra tuning is performed for the Monte-Carlo simulations of the bi-stable potential. A symmetric bi-stable potential is observed and a close match between Monte Carlo simulations and experiments is seen (see Figure~\ref{fig:550nm_multiplex}), validating the proposed simulation model.   

\section{Single Bit Memory, Information Erasure, Landauer's Principle}

A single bit memory has two discrete states - one or zero. A Brownian particle in a symmetric bi-stable potential is a commonly used model for a single bit memory. If the particle resides in the left well, we designate it the state zero and if it resides in the right well, we designate it the state one of the memory. The barrier height should be sufficient (higher than a few $k_BTs$) to ensure that particle does not cross over the barrier due to thermal fluctuations resulting in  the spontaneous loss of stored information. The depth of the two wells should be identical to ensure that the memory has no bias toward storing a zero or a one. Thus, it is equally likely the state of a memory bit, $M$, is a one or a zero, that is, $P(M=0)=P(M=1)=0.5$.

\begin{figure}
	\centering
	\includegraphics[scale = 0.35]{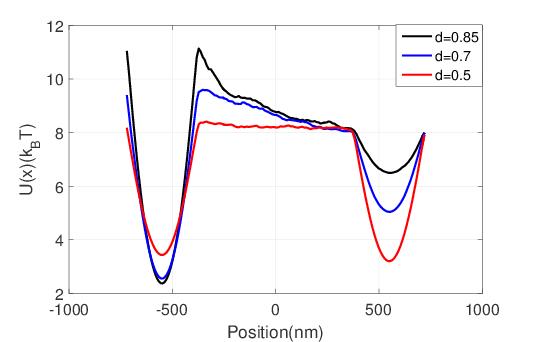} \vspace{-0.3 cm}
	\caption{The bi-stable potentials obtained using the Monte Carlo simulation framework described in the previous section for a variety of duty ratios. Higher the duty ratio, higher is the asymmetry.  
	}
	\label{fig:duty_sim}
	\vspace{-0.4 cm}
\end{figure}

Erasure of information stored can be achieved by the reset to zero operation. Irrespective of the stored value in a single bit memory, after undergoing the erasure process the final state is set to zero, that is, $M=0$. Thus, after erasure, $P(M=0) = 1, P(M=1)=0$. With regards to the bi-stable potential model of a memory bit, the Brownian particle needs to be transported to the left well irrespective of the initial location. This is achieved by multiplexing the laser with $d> 0.5$ between $-L$ and $L$ alternately, which results in an asymmetric bi-stable potential as shown in Figure \ref{fig:duty_sim}. It is seen that unequal duty ratio results in lowering the left well while lifting the right well, thus favoring the transport of the particle to the left from the right well (by lowering the barrier required to overcome) if it was initially in the right well or ensuring it stays in the left well (by increasing the barrier height of transition into the other well) if it was initially in the left well. 
 
Erasure process results in decrease in thermodynamic entropy of the system (the Brownian particle in a symmetric bi-stable well) as it results in reduction of the phase space volume for the Brownian particle of interest. Before erasure, the particle can cover twice the space than after undergoing erasure process. The decrease in thermodynamic entropy of the system  for the reset operation is $k_b\ln{2}$ and is accompanied by an average dissipation, $\langle Q_d \rangle$, of at least $k_BT\ln{2}$ amount of energy. Landauer's principle, states that erasure of information is necessarily dissipative and is quantified as,
\begin{align}
    \langle Q_d \rangle \geq k_BT \ln 2.
\end{align}
The decrease in thermodynamic entropy of the system ($k_B\ln 2$) \cite{shizume1995heat}, needs to be matched by at least $k_BT \ln 2$ of average dissipation to respect the $2^{nd}$ Law of Thermodynamics. The equality holds in the above inequality, if the erasure
process is performed in a quasi static manner. It is worth mentioning that, during the erasure, there is no average change in potential energy of the Brownian article, as the initial and final states are identical with respect to potential energy considerations.

\begin{figure}
	\centering
	\includegraphics[scale = 0.3]{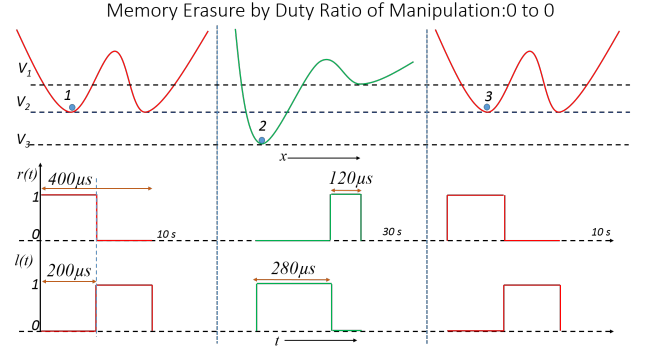} \vspace{-0.3 cm}
	\caption{Visual illustration of duty ratio protocol for bead initially in the left well. Here, $l(t)$ and $r(t)$ are binary signals which denote the status of laser at $-L$ and $L$ respectively, with the value $1$ for presence and $0$ for absence.  
	}
	\label{fig:020}
	\vspace{-0.4 cm}
\end{figure}

\begin{figure}
	\centering
	\includegraphics[scale = 0.3]{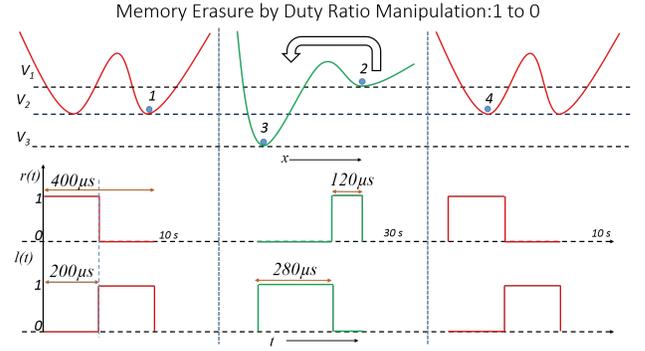} \vspace{-0.3 cm}
	\caption{Visual illustration of duty ratio protocol for bead initially in the right well. Here, $l(t)$ and $r(t)$ are binary signals which denote the status of laser at $-L$ and $L$ respectively, with the value $1$ for presence and $0$ for absence. 
	}
	\label{fig:120}
	\vspace{-0.4 cm}
\end{figure}

The steps below describe an erasure protocol using manipulation of duty ratio  in an optical tweezer setup (see Figure \ref{fig:020} and Figure \ref{fig:120}):
\begin{enumerate}
    \item Set duty ratio, $d=0.5$, for $10$ seconds to realize a single bit memory, with initial stored value zero or one.
    \item Set duty ratio, $d > 0.5$ for $\tau$ seconds to execute the erasure process (ensures the particle is transported to the left well).
    \item Set duty ratio, $d=0.5$ for $10$ seconds to get back to the model of single bit memory with stored value zero. 
\end{enumerate}
The choice of $\tau$ is such that it is in multiples of exit time from the right to the left well but lower than the exit time from left well to right one. The above mentioned protocol with $d > 0.7$ results in erasures with more than $95\%$ accuracy. For further details about the erasure protocol please refer to \cite{talukdar2017memory}.  

\section{Stochastic Thermodynamics}

In this section we briefly review the thermodynamics of Langevin systems as described in \cite{sekimoto2010stochastic}. According to the $1^{st}$ Law of Thermodynamics, $\langle \Delta U \rangle = \langle Q_d\rangle - \langle W \rangle$, where $W$ is the work done on the system and $dU$ is the change in potential energy. In an erasure process, as mentioned previously, $\langle dU \rangle = 0$, implying,
\begin{align}\label{eqn:first}
    \langle Q_d \rangle = \langle W\rangle.
\end{align}
The work done on the bead in a realization of erasure process by changing the duty-ratio (which results in modifying the potential felt by the bead), $W$, is given by \cite{sekimoto2010stochastic}, 
\begin{align}\label{eqn: work}
    W =  \sum_{j}[U(x(t_j)) - U(x(t_j))], 
\end{align}
where, $t_j$ denotes the time instants when the duty ratio is changed, and $t_j^-$ and $t_j^+$ denote the instants just before and after changing the duty ratio. Note that the potential $U(x)$ for each choice of duty ratio is obtained from the position data of the bead using (\ref{eqn: potential}). 
The average heat dissipation, which is equal to the average work done on the particle, obtained from 100 Monte Carlo realizations of erasure with a duty ratio of $0.7, 0.75, 0.8,0.85$ is shown in Figure \ref{fig:comparison}. 

\section{Optimal Erasures} \label{sec: Optimal Erasures}


In \cite{aurell2012refined}, the authors derive a  ``refined second law of thermodynamics'' for processes satisfying an overdamped Langevin equation, which presented in terms of the Wasserstein metric on probability distributions. 

The Wasserstein cost associated to probability measures, $\mu$ and $\nu$, is defined as
\[
    W_2^2(\mu, \nu) = \inf_{\pi \in \Pi(\mu,\nu)} \int_{\mathbb{R}^2} |x-y|^2 d \pi(x,y).
\]
where the infimum is taken over all $\pi$ couplings of the measures $\mu$ and $\nu$ on the real line. Explicitly, $\pi$ is a measure on $\mathbb{R}^2$ such that for Borel sets $A$ and $B$,
$
    \pi(A \times \mathbb{R}) = \mu(A), \hspace{3mm} \pi(\mathbb{R} \times B) = \nu(B)
$

The authors of \cite{aurell2012refined} derive
\begin{align}\label{eq: refined second law}
    \langle Q_d \rangle \geq \langle \Delta  H \rangle   + \gamma \frac{W^2_2( f_0, f_{t_f} )}{t_f},
\end{align}
where $\Delta H$ denotes the Hemholtz free energy, and $f_t$ stands for probability distribution satisfying the Fokker-Planck equation associated to \eqref{eqn: discrete_langevin}, with initial distribution $f_0$ and final distribution $f_{t_f}$. 

Applied to the current framework, (see \cite{talukdar2018analyzing} for more background) the refined second law \eqref{eq: refined second law} implies the following refinement of Landauer's bound,
\begin{align} \label{eq: refined Landauer}
    \langle Q_d \rangle \geq k_B T \ln (2)  + \gamma \frac{W^2_2( f_0, f_{t_f} )}{t_f}
\end{align}
where $f_0$ and $f_{t_f}$ denote the probability distributions associated to the energy landscapes before and after erasure. 
$W_2$ induces a metric on the space of probability distributions\footnote{Atleast those satisfying a technical integrability condition} whose numerical computation will be described below.

\begin{figure}[h] 
	\centering
	\includegraphics[scale = 0.55]{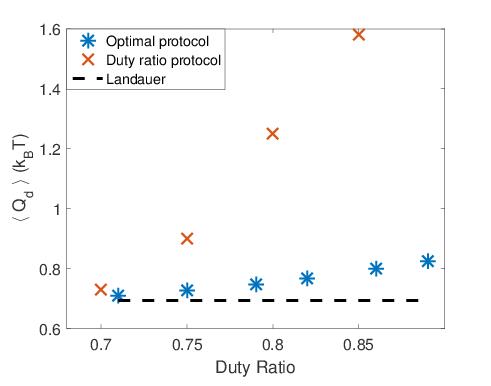} \vspace{-0.3 cm}
	\caption{Average heat dissipation in duty ratio protocol and optimal protocol.}
	\label{fig:comparison}
	\vspace{-0.4 cm}
\end{figure}

Though the theoretical existence and uniqueness of $W_2$ optimal couplings holds in very general settings \cite{villani2008optimal}, explicit formulas for determining couplings are much more elusive.  However, on the real line, the optimal coupling has a simple algorithmically accessible form. The trajectories of particles in an optimal transport are non-crossing, it follows that the optimal coupling $\pi$ must be order preserving. 
In the continuous setting this implies that the optimal coupling preserves quantiles, in the sense that when $F(x) = \mu(-\infty,x]$ and $G(y) = \nu (-\infty, y]$ the optimal is given by 
   \[
        \pi(A,B) = \mu(A \cap T^{-1}(B))
   \]
    when $T = G^{-1}F$.  This implies that 
    \begin{align} \label{eqn: Wasserstein Map}
        W_2^2(\mu, \nu) 
            &= \int_\mathbb{R} |x-Tx|^2 d \mu(x) 
                    \\
            &= \int_\mathbb{R} |x - G^{-1}F(x)|^2 d F(x)
    \end{align}
We employ the above relationship and its numerical approximation to compute the optimal cost between the initial and the final probability distribution functions for the position of the bead resulting from the Monte Carlo simulations  of the duty ratio based protocol. This is compared with the heat dissipated  by the duty ratio based protocol  as computed using (\ref{eqn: work}) corresponding to the same initial pdf  and  the same final pdf of locations of the bead.  The time to complete the protocol is fixed at $t_f=30.$  The average heat dissipation associated with the optimal protocol for various values of duty ratio is shown in Figure \ref{fig:comparison}. It is seen that for the duty ratio protocol is close to the optimal protocol with regard to heat dissipation for a duty ratio around 0.7. However, as the duty ratio increases, the gap between the average heat dissipation obtained using the duty ratio protocol and the optimal protocol increases. Moreover, the average heat dissipation associated with the optimal protocol increases as the duty ratio increases and is close to the Landauer bound for a duty ratio of around $0.7$. 



\section{Finite-Time Protocols}
In the earlier section we computed the Wasserstein metric of a initial bead location pdf and final bead location pdf and compared it to the heat dissipated by a specific protocol (duty ratio protocol). Here we focus on providing guidance on how the transfer of pdf's can be achieved to bring the implemented protocol to have heat dissipated closer to the optimal Wasserstein based cost.  
The geometry of Wasserstein space, and simple properties of the Hemholtz free energy suggest that erasure protocols could reduce heat dissipation by transporting a particle in such a way that its distribution in time $t$ will approximate Wasserstein geodesics in $W_2$. 

Such geodesics can be recovered from optimal transportion plans.  Indeed, when $T$ gives an optimal transportation map from $f_0$ to $f_1$, then 
\begin{equation} \label{eqn: displacement interpolation}
    T_t = (1-t) I + t T
\end{equation} gives a unit speed ``displacement interpolation'' of of $f_0$ an $f_1$ along a Wasserstein geodesic.  

As depicted in the figure below the protocol of \cite{talukdar2017memory} is sub-optimal for regimes with duty ratio $>.75$.  We hypothesize that a protocol that encourages the bead's distribution to mimic that transportation depicted in figure 10, can reduce heat dissipation.

\section{Results}

As demonstrated in the figure below, the cost for an optimal protocol erasure and the cost realized in the experiments of \cite{bhaban2016noise, talukdar2017memory} deviate only very slightly in lower duty ratio regimes (around $d=0.7$), while in high duty ratio regimes, the cost could be cut in half by an optimal protocol. 

Figure 10 depicts an optimal transport geodesic between duty ratio $.75$ and $.5$.  It is achieved by computing the optimal transport map $T = G^{-1} F$ from \eqref{eqn: Wasserstein Map}, as described in (\ref{sec: Optimal Erasures}).  The probability distribution at a time $t$ is given by the pushforward of the initial distribution $f_0$ by the map $T_{t/t_f}$ as described in \eqref{eqn: displacement interpolation}. 

\begin{figure}[h] \label{fig: displacement inter}
	\centering
	\includegraphics[scale = 0.25]{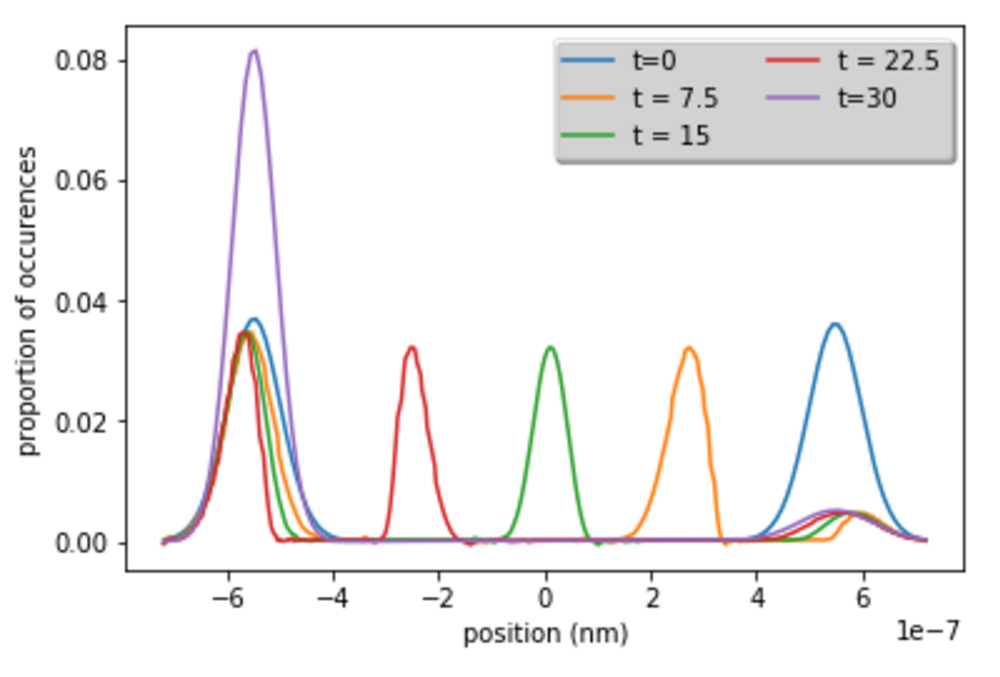} \vspace{-0.3 cm}
	\caption{Wasserstein optimal path for erasure in 30 seconds. Purple curve is the final distribution and blue curve is the initial distribution. The mass from the right is progressively moved toward the left.}
	\vspace{-0.4 cm}
\end{figure}

\vspace{3mm}

\section{CONCLUSIONS AND FUTURE WORKS}

Recent theoretical work has given a time and space dependent sharpening of Landauer's inequality, which we have shown to be relevant to optical tweezer realizations of an erasure.  We have demonstrated some experimental agreement with these theories, and hypothesized that an erasure protocol, that sends a particle along a Wasserstein geodesic should reduce heat dissipation.  It would be of immediate significance to design and implement a tweezer protocol to test these ideas.

\section{ACKNOWLEDGMENTS}

The authors thank Prof. Tryphon Georgiou, University of California, Irvine and Prof. Yongxin Chen, Iowa State University, for directing our attention toward optimal erasure protocols. We are also thankful to Mr. Shreyas Bhaban at the University of Minnesota for his efforts toward understanding and developing the duty ratio protocol.

\bibliographystyle{plain}
\bibliography{bibibi}

\end{document}